\documentclass[preprintnumbers,showkeys,amsmath,amssymb]{revtex4}
\usepackage{graphicx}
\usepackage{dcolumn}
\usepackage{bm}
\usepackage{epstopdf}

\begin{document}

\title{NOON state generation with phonons in acoustic wave resonators assisted by a nitrogen-vacancy-center ensemble}

\author{Jiu-Ming Li,
 Ming Hua\footnote{Corresponding
author:huaming@tjpu.edu.cn}, and Xue-Qun Yan }

\address{ Department of Applied Physics, School of Physical Science and Technology, Tianjin Polytechnic University, Tianjin 300387, China}

\begin{abstract}
Since the quality factor of an acoustic wave resonator (AWR) reached $10^{11}$, AWRs have been regarded as a good carrier of quantum information. In this paper, we propose a scheme to construct a NOON state with two AWRs assisted by a nitrogen-vacancy-center ensemble (NVE). The two AWRs cross each other vertically, and the NVE is located at the center of the crossing. By considering the decoherence of the system and using resonant interactions between the AWRs and the NVE, and the single-qubit operation of the NVE, a NOON state can be achieved with a fidelity higher than $98.8\%$ when the number of phonons in the AWR is $N \le 3$.
\end{abstract}

\keywords{acoustic wave resonator; phonon; nitrogen-vacancy-center ensemble; quantum entanglement; quantum electrodynamics}

\maketitle

\section{INTRODUCTION}

Quantum entanglement is not only one of the significant features of quantum mechanics but also an important application in quantum information processing (QIP) \cite{QE1,QE2,r1,r2,r3,r4,r5,r6,r7,r8,r9,r10,r13,r14}. Over the past decades, many works have been proposed to creat entanglements using atoms or photons assisted by various quantum systems \cite{NORI4,NORI7}, such as cavity quantum electrodynamics (cavity-QED) \cite{cQED1,YCP1,LPB1,XZY,ZS,LGL1,Nori1}, circuit quantum electrodynamics (circuit-QED) \cite{NORI5,NORI6,NORI10,cir1,cir2,cir3,TL1,LPB2,LFL1,Nori2,YCP2,LFL2},trapped ions \cite{ti1,ti2,ti3,ti4}, quantum dots \cite{LGL3,QD1,QD2,QD3,QD4,QD5}, cold atoms \cite{ca1,ca2}, nitrogen-vacancy (NV) centers \cite{ni1,ni3,ni4,ni5,WHF}, nuclear magnetic resonance \cite{NMR,Long1,NHQCLong,LGL2,r12}, acoustic wave resonators (AWRs) \cite{AWR}, optomechanical systems \cite{TL2,os,Nori5,r11}, and atomic ensembles in free space \cite{AE}.

In recent years, acoustic wave resonators, whose quality-factor (Q-factor) has been increased to $\sim10^{11}$ \cite{IEEE}, have attracted more and more attention for QIP \cite{SAW1,SAW2,SAW3,SAW4,OL,SAW5,SAW6,APL,SAW7,SAW8,SAW9,Kervinen}. For example, in 2012, Goryachev \emph{et al}. \cite{APL} measured a quartz bulk acoustic wave resonator and showed that the Q-factor of the resonator can be continually increased to $10^9$. In 2018, Cai \emph{et al}. \cite{OL} proposed a scheme to achieve a single phonon source based on the nonlinearity generated by the four-level NV-centers in a diamond photonic crystal resonator. In 2017, Noguchi \emph{et al}. \cite{SAW6} demonstrated an ultrasensitive measurement of fluctuations in an AWR and up-converted the excitation in the AWR to an excitation in a microwave resonator in a hybrid system consisting of an AWR, a microwave resonator, and a superconducting qubit.

To achieve the QIP with AWRs, many works have studied the coupling between a nonlinear quantum system and an AWR both experimentally and theoretically. Examples include the coupling between an AWR and a superconducting qubit \cite{ssq1,ssq2,ssq3,qa,ssq4,NORI3}, an NV-center \cite{SAW3,NORI1,NORI2}, the phononic QED \cite{SAW2,pQ}, and the quantum dots \cite{aqd1,aqd2,aqd3}.
Among these couplings, due to the long coherence time \cite{ct,Nori3} and good manipulability of the NV-centers \cite{YZQ1,LGL4,YZQ2,Nori4}, the coupling between NV-centers and the AWRs has attracted much attention in recent years. For example, in 2016, Golter \emph{et al}. \cite{SAW3} realized the strong coupling between a nitrogen-vacancy-center ensemble (NVE) and an AWR, and they also realized the coherent population trapping and optically driven spin transitions. In the same year \cite{SAW4}, they demonstrated the quantum control of an NV-center in the resolved-sideband regime by coupling the NV-center to the optical and the AWR fields.

In this paper, we propose a simple scheme to generate a NOON state with AWRs in a system consisting of two AWRs coupled to an NVE for the first time.
The operations used here are resonant interactions between the AWRs (microwave pulse) and the NVE, which help us to achieve a high fidelity NOON state in a short time.
To demonstrate the feasibility of the scheme, we numerically simulate the fidelities of the NOON states with the AWR phonon number as $N=1$, $2$, and $3$ by considering the decoherence of the system, and the fidelities of which reach $99.54\%$, $99.18\%$, and $98.80\%$, respectively.

\section{GENERATION OF NOON STATE WITH THE AWRS}

\begin{figure}[!htbp]       
\centering
\includegraphics[width=8.5cm,angle=0]{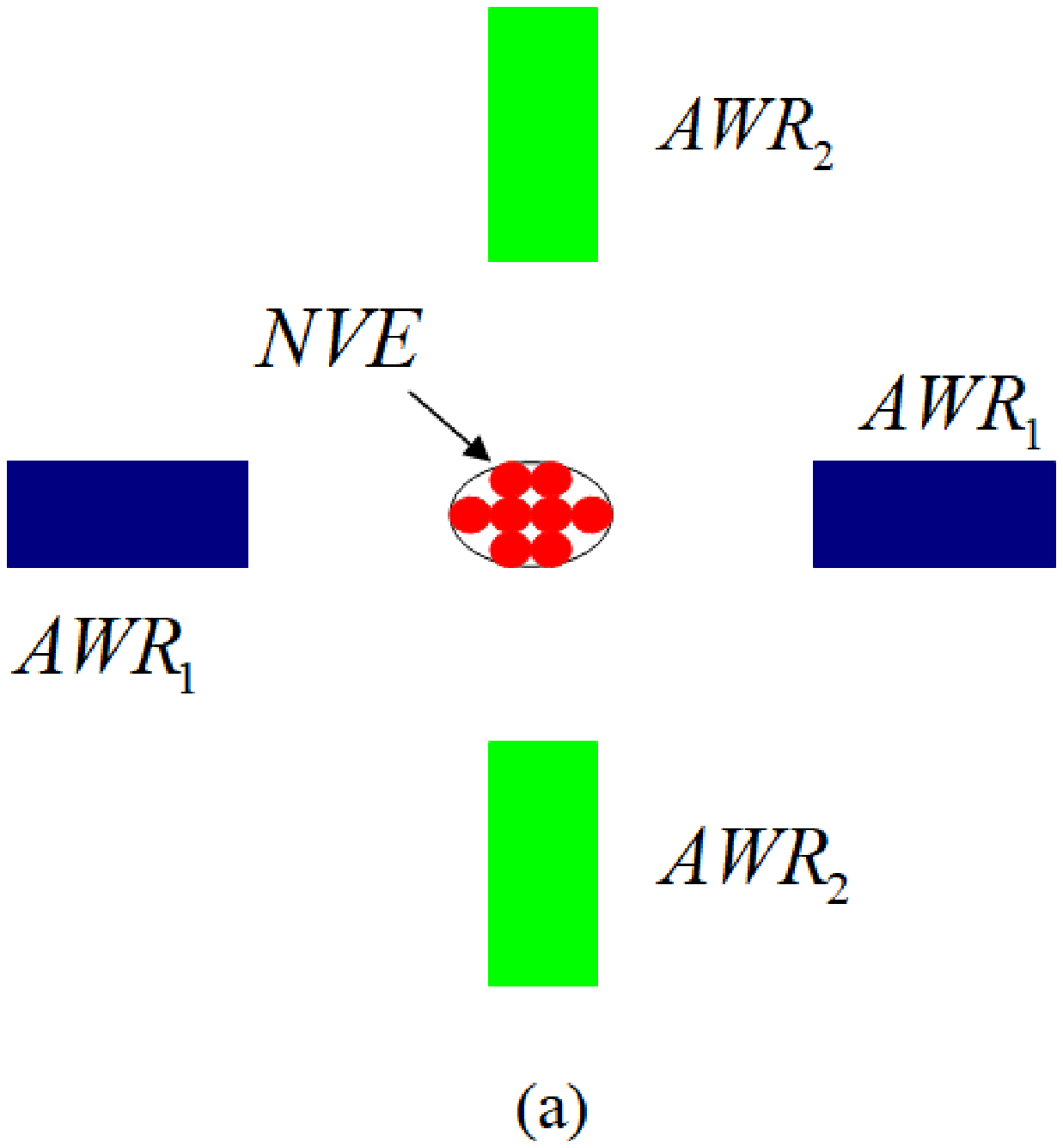}\!\!\!\!\!\!\!\!\!\!\!\!\!\!\!\!\!\!\!\!\!\!\!\!\!\!\!\!\!\!\!\!\!
\includegraphics[width=8.5cm,angle=0]{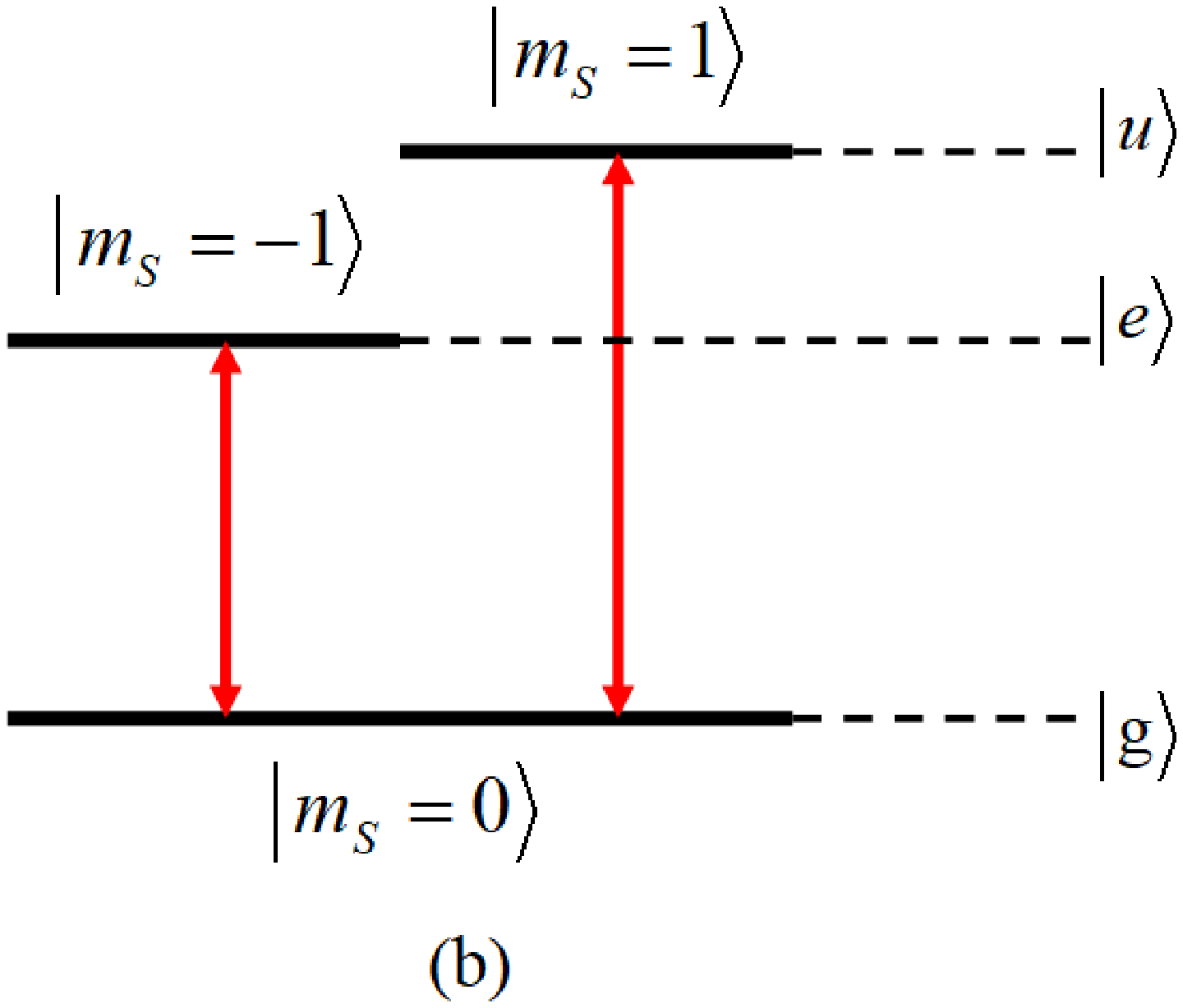}
\caption{(Color online) (a) Setup of the system consisting of two AWRs coupled to an NVE. (b) Energy levels of an NV-center under an external field $\vec{B}$. $\left|m_s=0\right\rangle$, $\left|m_s=-1\right\rangle$, and $\left|m_s=1\right\rangle$ are labelled as $\left|g\right\rangle$, $\left|e\right\rangle$, and $\left|u\right\rangle$, respectively. A microwave pulse with strength $\Omega_e$ ($\Omega_u$) is applied to flip the states $\left|g\right\rangle$ and $\left|e\right\rangle$ ($\left|g\right\rangle\leftrightarrow\left|u\right\rangle$) of the NVE.}\label{fig1}
\end{figure}

To construct a NOON state with AMRs, we consider a system consisting of two AWRs coupled to an NVE as shown in Fig.\ref{fig1}(a). The two AWRs cross each other vertically, and the NVE is located at the center of the crossing. The electronic ground state of the NV-center has a spin of $S=1$ with an energy splitting between the states $\left|m_s=0\right\rangle$ and $\left|m_s=\pm1\right\rangle$ with  frequency $\omega_{\pm}\approx2\pi\times2.88$ GHz in a zero magnetic field. By applying an external magnetic field $\vec{B}$ parallel to the axis between the nitrogen and the vacancy, the states $\left|m_s=\pm1\right\rangle$ can be split. For simplicity, the states of the NV-centers $\left|m_s=0\right\rangle$, $\left|m_s=-1\right\rangle$, and $\left|m_s=1\right\rangle$ are labelled as $\left|g\right\rangle$, $\left|e\right\rangle$, and $\left|u\right\rangle$, respectively, with where the energies are characterized by $E_g<E_e$ and $E_g<E_u$. $\omega_{ge}$ ($\omega_{gu}$) is defined as the transition frequency between states $\left|g\right\rangle$ and $\left|e\right\rangle$ ($\left|g\right\rangle$ and $\left|u\right\rangle$). Here, $\omega_{ge}$ can be tuned by $\vec{B}$, and $\omega_{gu}$ is kept unchanged as $2\pi\times2.88$ GHz. The states $\left|g\right\rangle$ and $\left|e\right\rangle(\left|g\right\rangle$ and $\left|u\right\rangle)$ of the NVE can be flipped by applying a microwave pulse with frequency $\omega_{e}$ ($\omega_{u}$) and strength $\Omega_e$ ($\Omega_u$) as shown in Fig.\ref{fig1}(b). In addition, the frequencies of the AWRs and the transition $|g\rangle \leftrightarrow |u\rangle$ of the NVE should be far detuned from each other largely.

There are two stages with ($2N+2M+3$) steps to generate the NOON state. First, we focus on the first stage which contains ($2N+1$) steps. In this stage, the two transitions $\left|g\right\rangle\leftrightarrow\left|e\right\rangle$ and $\left|g\right\rangle\leftrightarrow\left|u\right\rangle$ of the NVE are set to be far detuned from AWR$_2$ largely all the time. The initial state of the whole system should be prepared as
\begin{eqnarray}       
\left|\psi\right\rangle_I &=& \left|g\right\rangle\left|0\right\rangle_1\left|0\right\rangle_2.
\end{eqnarray}
Here, the subscript 1(2) represents the AWR$_{1(2)}$.

The operations of the first stage containing $(2N+1)$ steps are described as follows:

Step $1$: A microwave pulse with strength $\Omega_e$ is applied to resonate with the transition $\left|g\right\rangle\leftrightarrow\left|e\right\rangle$ of the NVE to form the Hamiltonian
$H_e=\hbar\Omega_e\left(\sigma_{ge}^++H.c.\right)$,
where $\sigma_{ge}^+$ represents the raising operator of the transition $|g\rangle\leftrightarrow|e\rangle$. Here, the two transitions of the NVE should be far detuned from AWR$_1$ and AWR$_2$ largely. Then, the state of the system will evolve from $\left|\psi\right\rangle_I$ to
\begin{eqnarray}         
\left|\psi\right\rangle_1 &=& \frac{1}{\sqrt{2}}\left(\left|g\right\rangle-i\left|e\right\rangle\right)\left|0\right\rangle_1\left|0\right\rangle_2
\end{eqnarray}
after an operation time $t=\pi/\left(4\Omega_e\right)$.

Step $2$: A microwave pulse with strength $\Omega_u$ is applied to resonate with the transition $\left|g\right\rangle\leftrightarrow\left|u\right\rangle$ of the NVE when the two transitions of the NVE are detuned from AWR$_1$ and AWR$_2$.
After an operation time of $t=\pi/\left(2\Omega_u\right)$, state $\left|g\right\rangle$ of the NVE is excited to $\left|u\right\rangle$ with a $-i$ phase shift, and the state of the whole system becomes
\begin{eqnarray}        
\left|\psi\right\rangle_2 &=& \frac{-i}{\sqrt{2}}\left(\left|e\right\rangle+\left|u\right\rangle\right)\left|0\right\rangle_1\left|0\right\rangle_2.
\end{eqnarray}

Step $3$: By letting the transition $|g\rangle \leftrightarrow |e\rangle$ of the NVE resonate with AWR$_1$ ($\omega_{ge}=\omega_1$), state $|\psi\rangle_2$ will evolve to
\begin{eqnarray}        
\left|\psi\right\rangle_3 &=& \frac{-i}{\sqrt{2}}\left(-i\left|g\right\rangle\left|1\right\rangle_1+\left|u\right\rangle\left|0\right\rangle_1\right)
\left|0\right\rangle_2
\end{eqnarray}
after an operation time $t=\pi/\left(2g_1^{ge}\right)$. Here, $\omega_{1(2)}$ is the frequency of the AWR$_{1(2)}$. $g_1^{ge}$ is the coupling strength between the AWR$_1$ and the transition $\left|g\right\rangle\leftrightarrow\left|e\right\rangle$ of the NVE.

Step $4$: A microwave pulse with strength $\Omega_e$ is applied to excite the state $\left|g\right\rangle$ to $\left|e\right\rangle$ after an operation time of $t=\pi/\left(2\Omega_e\right)$. Then, the state of the system evolves from $|\psi\rangle_3$ to
\begin{eqnarray}         
\left|\psi\right\rangle_4 &=& \frac{-i}{\sqrt{2}}\left[-i\left(-i\right)\left|e\right\rangle\left|1\right\rangle_1+\left|u\right\rangle\left|0\right\rangle_1\right]
\left|0\right\rangle_2.
\end{eqnarray}

Step $5$: As in the step $3$, we let the transition $\left|g\right\rangle\leftrightarrow\left|e\right\rangle$ of the NVE resonate with the AWR$_1$. After an interaction time $t=\pi/\left(2g_1^{ge}\right)$, the state of the system evolves from $\left|\psi\right\rangle_4$ to
\begin{eqnarray}         
\left|\psi\right\rangle_5 &=& \frac{-i}{\sqrt{2}}\left[-i\left(-1\right)\left|g\right\rangle\left|2\right\rangle_1+\left|u\right\rangle\left|0\right\rangle_1\right]
\left|0\right\rangle_2.
\end{eqnarray}

Step $j$ ($j=6,7,...,2N+1$): Repeating steps $4$ and $5$ successively with $N+1$ times, the state of the system will evolve to
\begin{eqnarray}         
\left|\psi\right\rangle_{2N+1} &=& \frac{-i}{\sqrt{2}}\left[-i\left(-1\right)^{N-1}\left|g\right\rangle\left|N\right\rangle_1
+\left|u\right\rangle\left|0\right\rangle_1\right]\left|0\right\rangle_2.
\end{eqnarray}

Then, by tuning the two transitions $\left|g\right\rangle\leftrightarrow\left|e\right\rangle$ and $\left|g\right\rangle\leftrightarrow\left|u\right\rangle$ of the NVE to be far detuned from AWR$_1$ largely, we will give the operations of ($2M+2$) steps of the second stage to achieve the NOON state as follows:

Step $2N+2$: A microwave pulse with strength $\Omega_u$ is applied to flip the states $\left|g\right\rangle$ and $\left|u\right\rangle$ of the NVE. After an operation time $t=\pi/\left(2\Omega_u\right)$, the state of the system becomes
\begin{eqnarray}          
\left|\psi\right\rangle_1^{\prime} &=& \frac{-1}{\sqrt{2}}\left[-i\left(-1\right)^{N-1}
\left|u\right\rangle\!\left|N\right\rangle_1\left|0\right\rangle_2+\left|g\right\rangle\left|0\right\rangle_1\left|0\right\rangle_2\right].
\end{eqnarray}

Step $2N+3$: A microwave pulse with strength $\Omega_e$ is applied to excite the state $\left|g\right\rangle$ to $\left|e\right\rangle$. State $\left|\psi\right\rangle_1^{\!\prime}$ will evolve to
\begin{eqnarray}          
|\psi\rangle_2^{\prime} &=& \frac{-1}{\sqrt{2}}\left[-i\left(-1\right)^{N\!-\!1}|u\rangle\left|N\right\rangle_1\left|0\right\rangle_2
+\left(-i\right)\left|e\right\rangle\left|0\right\rangle_1\left|0\right\rangle_2\right].
\end{eqnarray}

Step $2N+4$: Resonating the AWR$_2$ with the transition $\left|g\right\rangle\leftrightarrow\left|e\right\rangle$ of the NVE ($\omega_{ge}=\omega_2$), the state of the system will evolve from $|\psi\rangle_2'$ to
\begin{eqnarray}           
|\psi\rangle_3^{\prime} &=& \frac{-1}{\sqrt{2}}\left[-i\left(-1\right)^{N\!-\!1}\left|u\right\rangle\left|N\right\rangle_1\left|0\right\rangle_2
+\left(-1\right)\left|g\right\rangle\left|0\right\rangle_1\left|1\right\rangle_2\right]
\end{eqnarray}
after an operation time of $t=\pi/\left(2g_2^{ge}\right)$. Here, $g_2^{ge}$ is the coupling strength between the AWR$_2$ and the transition $\left|g\right\rangle\leftrightarrow\left|e\right\rangle$ of the NVE.

Then, repeating the operations of steps $2N+3$ and $2N+4$ with $M-1$ times, the state of the whole system evolves to
\begin{eqnarray}            
\left|\psi\right\rangle_{2M\!-\!1}^{\prime} &=& \frac{-1}{\sqrt{2}}\bigg[-i\left(-1\right)^{N-1}\left|u\right\rangle\left|N\right\rangle_1
\left|0\right\rangle_2+\left(-1\right)^{M-1}\left|g\right\rangle\left|0\right\rangle_1\left|M\!-\!1\right\rangle_2\bigg].
\end{eqnarray}

Step $2N+2M+1$: Applying a microwave pulse with strength $\Omega_e$ to excite the state $\left|g\right\rangle$ to the state $\left|e\right\rangle$, state $\left|\psi\right\rangle_{2M\!-\!1}^{\prime}$ will evolve to
\begin{eqnarray}             
\left|\psi\right\rangle_{2M}^{\prime} &=& \frac{-1}{\sqrt{2}}\bigg[-i(-1)^{N\!-\!1}\left|u\right\rangle\left|N\right\rangle_1
\left|0\right\rangle_2+\left(-i\right)(-1)^{M\!-\!1}\left|e\right\rangle\left|0\right\rangle_1\left|M-1\right\rangle_2\bigg].
\end{eqnarray}

Step $2N+2M+2$: The same as the step $2N+2$. The state of the system evolves from $\left|\psi\right\rangle_{2M}^{\prime}$ to
\begin{eqnarray}             
\left|\psi\right\rangle_{2M\!+\!1}^{\prime} &=& \frac{-1}{\sqrt{2}}\bigg[\left(-1\right)^{N}\left|g\right\rangle\left|N\right\rangle_1
\left|0\right\rangle_2+\left(-i\right)\left(-1\right)^{M\!-\!1}\left|e\right\rangle\left|0\right\rangle_1\left|M-1\right\rangle_2\bigg].
\end{eqnarray}

Finally, resonating the transition $\left|g\right\rangle\leftrightarrow\left|e\right\rangle$ of the NVE with the AWR$_2$, the state of the system becomes
\begin{eqnarray}             
\left|\psi\right\rangle_f &=& \frac{-1}{\sqrt{2}}\left[(-1)^N\left|N\right\rangle_1\left|0\right\rangle_2+
(-1)^M\left|0\right\rangle_1\left|M\right\rangle_2\right]\left|g\right\rangle,
\end{eqnarray}
which is just the NOON state of the AWRs.

\section{NUMERICAL SIMULATION}

The operations for generating the NOON state with the AWRs contain four kinds of resonant interactions: the microwave pulse with strength $\Omega_e$ resonates with the transition $\left|g\right\rangle\leftrightarrow\left|e\right\rangle$ of the NVE, the microwave pulse with strength $\Omega_u$ resonates with the transition $\left|g\right\rangle\leftrightarrow\left|u\right\rangle$ of the NVE, the transition $\left|g\right\rangle\leftrightarrow\left|e\right\rangle$ of the NVE resonates with the AWR$_1$, and the transition $\left|g\right\rangle\leftrightarrow\left|e\right\rangle$ of the NVE resonates with the AWR$_2$. Hamiltonians of the system for these interactions are given below:

First, when the microwave pulse with strength $\Omega_e$ resonates with the transition $\left|g\right\rangle\leftrightarrow\left|e\right\rangle$ of the NVE, the Hamiltonian of the system can be written as
\begin{eqnarray}         
H_e^r &=& \hbar\Omega_e\left(\sigma_{ge}^++H.c.\right) + \hbar\Omega_e\left(\sigma_{gu}^+e^{i\Delta_e^u t}+H.c.\right) + \hbar g_1^{ge}\left(\sigma_{ge}^+b_1 e^{i\Delta_1^{ge}t}+H.c.\right) \nonumber\\
&&+ \hbar g_1^{gu}\left(\sigma_{gu}^+\,b_1 e^{i\Delta_1^{gu}t}+H.c.\right) + \hbar g_2^{ge}\left(\sigma_{ge}^+b_2 e^{i\Delta_2^{ge}t}+H.c.\right) + \hbar g_2^{gu}\left(\sigma_{gu}^+b_2 e^{i\Delta_2^{gu}t}+H.c.\right).
\end{eqnarray}
Here, $b_1$ ($b_2$) is the annihilation operator of the AWR$_1$ (AWR$_2$), and $\sigma_{ge}^+$ ($\sigma_{gu}^+$) represents the raising operator of the transition $\left|g\right\rangle\leftrightarrow\left|e\right\rangle$ ($\left|g\right\rangle\leftrightarrow\left|u\right\rangle$) of the NVE. $g_1^{gu}$ ($g_2^{gu}$) is the coupling strength between the AWR$_1$ (AWR$_2$) and the transition $\left|g\right\rangle\leftrightarrow\left|u\right\rangle$. $\Delta_1^{ge}=\omega_{ge}-\omega_1$, $\Delta_2^{ge}=\omega_{ge}-\omega_2$, $\Delta_1^{gu}=\omega_{gu}-\omega_1$, $\Delta_2^{gu}=\omega_{gu}-\omega_2$, and $\Delta_e^u=\omega_{gu}-\omega_e$. $\omega_e$ is the frequency of the microwave pulse with strength $\Omega_e$.

Second, the microwave pulse with strength $\Omega_u$ is applied to resonate with the transition $\left|g\right\rangle\leftrightarrow\left|u\right\rangle$ of the NVE. The Hamiltonian of the system can be expressed as
\begin{eqnarray}       
H_u^r &=& \hbar\Omega_u\left(\sigma_{gu}^++H.c.\right) +  \hbar\Omega_u\left(\sigma_{ge}^+e^{i\Delta_u^e t}+H.c.\right) + \hbar g_1^{ge}\left(\sigma_{ge}^+b_1 e^{i\Delta_1^{ge}t}+H.c.\right) \nonumber\\
&&+\hbar g_1^{gu}\left(\sigma_{gu}^+b_1 e^{i\Delta_1^{gu}t}+H.c.\right) + \hbar g_2^{ge}\left(\sigma_{ge}^+b_2 e^{i\Delta_2^{ge}t}+H.c.\right) + \hbar g_2^{gu}\left(\sigma_{gu}^+\,b_2 e^{i\Delta_2^{gu}t}+H.c.\right),
\end{eqnarray}
where $\Delta_u^e=\omega_{ge}-\omega_u$.

Third, when the frequency of the transition $\left|g\right\rangle\leftrightarrow\left|e\right\rangle$ of the NVE is tuned to resonate with AWR$_1$, the Hamiltonian of the system becomes
\begin{eqnarray}          
H_1^r &=& \hbar g_1^{ge} \! \left(\sigma_{ge}^+b_1 \!+\! H.c.\right) \!+\! \hbar g_1^{gu} \! \left(\sigma_{gu}^+b_1 e^{i\Delta_1^{gu}t} \!+\! H.c.\right) \!+\! \hbar g_2^{ge} \! \left(\sigma_{ge}^+b_2 e^{i\Delta_2^{ge}t} \!+\! H.c.\right) \!+\! \hbar g_2^{gu} \! \left(\sigma_{gu}^+b_2 e^{i\Delta_2^{gu}t} \!+\! H.c.\right).
\end{eqnarray}

\begin{table}[!hbp]     
\centering
\caption{Parameters for generating a NOON state with $N=M=1$.}\label{tab1}
\begin{tabular}{c|ccc}
\hline
\, step \,&\, $\Omega_e/2\pi$(MHz) \,& \,$\Omega_u/2\pi$(MHz)\,  &\, $\omega_{ge}/2\pi$(GHz)\, \\
\hline
(1) & 1.4 & 0 & 2.7 \\
(2) & 0 & 1.3 & 2.7 \\
(3) & 0 & 0 & 0.1525 \\
(4) & 0 & 1.3 & 2.7 \\
(5) & 0.9 & 0 & 2.7 \\
(6) & 0 & 0.9 & 2.7 \\
(7) & 0 & 0 & 0.1848 \\
\hline
\end{tabular}
\end{table}

\begin{table}[!hbp]     
\centering
\caption{Parameters for generating a NOON state with $N=M=2$.}\label{tab2}
\begin{tabular}{c|ccc}
\hline
\, step \,&\, $\Omega_e/2\pi$(MHz) \, & \,$\Omega_u/2\pi$(MHz)\, & \, $\omega_{ge}/2\pi$(GHz)\, \\
\hline
(1) & 1.4 & 0 & 2.7 \\
(2) & 0 & 1.3 & 2.7 \\
(3) & 0 & 0 & 0.1525 \\
(4) & 0.9 & 0 & 2.7 \\
(5) & 0 & 0 & 0.1525 \\
(6) & 0 & 1.3 & 2.7 \\
(7) & 0.9 & 0 & 2.7 \\
(8) & 0 & 0 & 0.1848 \\
(9) & 0.9 & 0 & 2.7 \\
(10) & 0 & 0.9 & 2.7 \\
(11) & 0 & 0 & 0.1848 \\
\hline
\end{tabular}
\end{table}

\begin{table}[!hbp]        
\centering
\caption{Parameters for generating a NOON state with $N=M=3$.}\label{tab3}
\begin{tabular}{c|ccc}
\hline
\, step \,&\, $\Omega_e/2\pi$(MHz) \, & \,$\Omega_u/2\pi$(MHz)\, & \, $\omega_{ge}/2\pi$(GHz)\, \\
\hline
(1) & 1.4 & 0 & 2.7  \\
(2) & 0 & 1.3 & 2.7  \\
(3) & 0 & 0 & 0.1525  \\
(4) & 0.9 & 0 & 2.7  \\
(5) & 0 & 0 & 0.1525  \\
(6) & 0.9 & 0 & 2.7  \\
(7) & 0 & 0 & 0.1525  \\
(8) & 0 & 1.3 & 2.7  \\
(9) & 0.9 & 0 & 2.7  \\
(10) & 0 & 0 & 0.1848  \\
(11) & 0.9 & 0 & 2.7  \\
(12) & 0 & 0 & 0.1848  \\
(13) & 0.9 & 0 & 2.7  \\
(14) & 0 & 0.9 & 2.7  \\
(15) & 0 & 0 & 0.1848  \\
\hline
\end{tabular}
\end{table}

In the last case, resonating the frequency of the transition $\left|g\right\rangle\leftrightarrow\left|e\right\rangle$ with the AWR$_2$ and making the AWR$_2$ far detuned from the transition $\left|g\right\rangle\leftrightarrow\left|u\right\rangle$ of the NVE largely, the Hamiltonian can be expressed as
\begin{eqnarray}       
H_2^r &=& \hbar g_2^{ge}\!\left(\sigma_{ge}^+b_2 \!+\! H.c.\right) \!+\! \hbar g_2^{gu} \! \left(\sigma_{gu}^+b_2 e^{i\Delta_2^{gu}t} \!+\! H.c.\right) \!+\! \hbar g_1^{ge} \!\left(\sigma_{ge}^+b_1 e^{i\Delta_1^{ge}t} \!+\! H.c.\right) \!+\! \hbar g_1^{gu} \! \left(\sigma_{gu}^+b_1 e^{i\Delta_1^{gu}t} \!+\! H.c.\right).
\end{eqnarray}

\begin{figure}[!htbp]      
\centering
\includegraphics[width=8.5cm,angle=0]{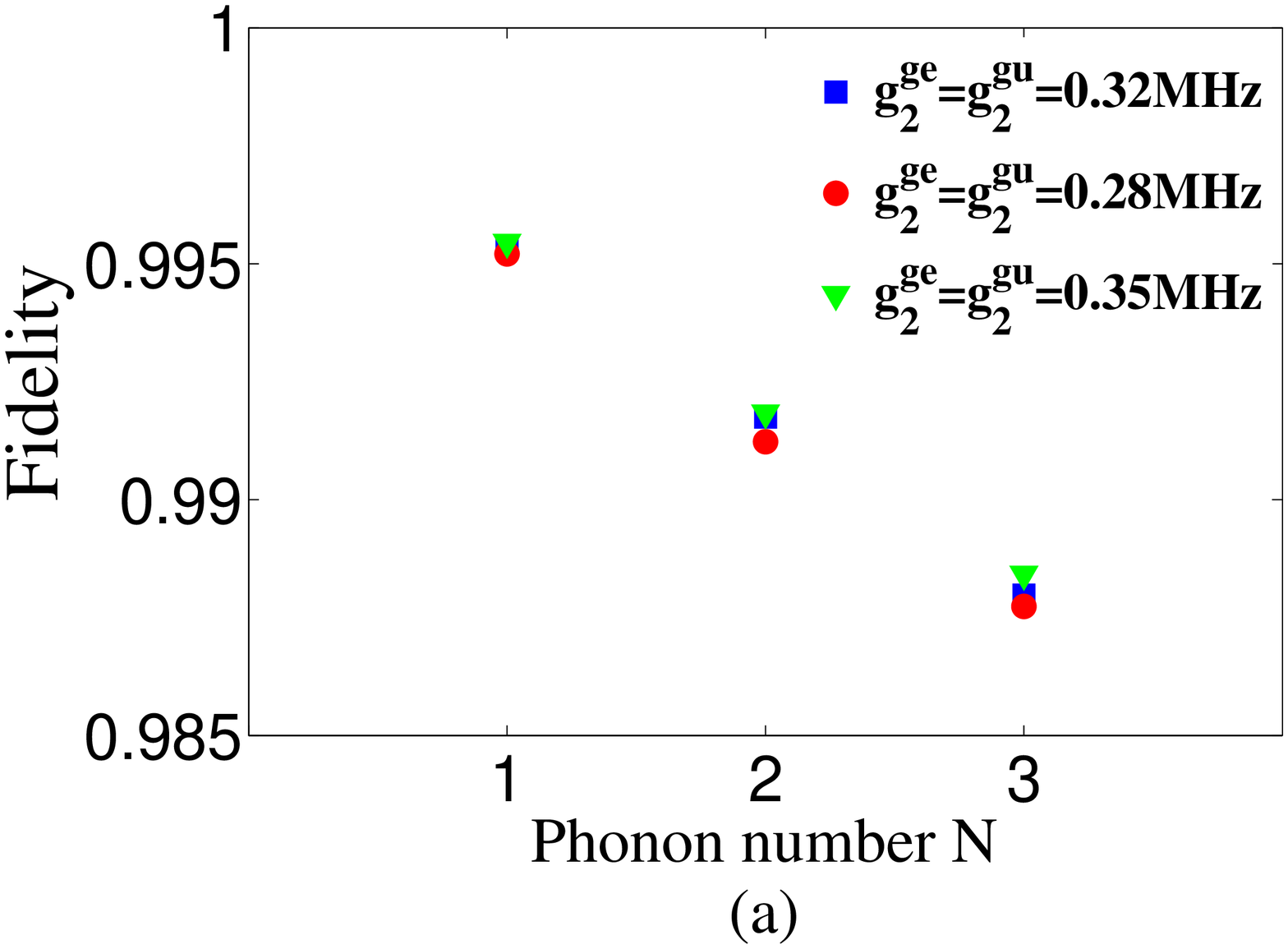}\!\!\!\!\!\!\!\!\!\!\!\!\!\!\!\!\!
\includegraphics[width=8.5cm,angle=0]{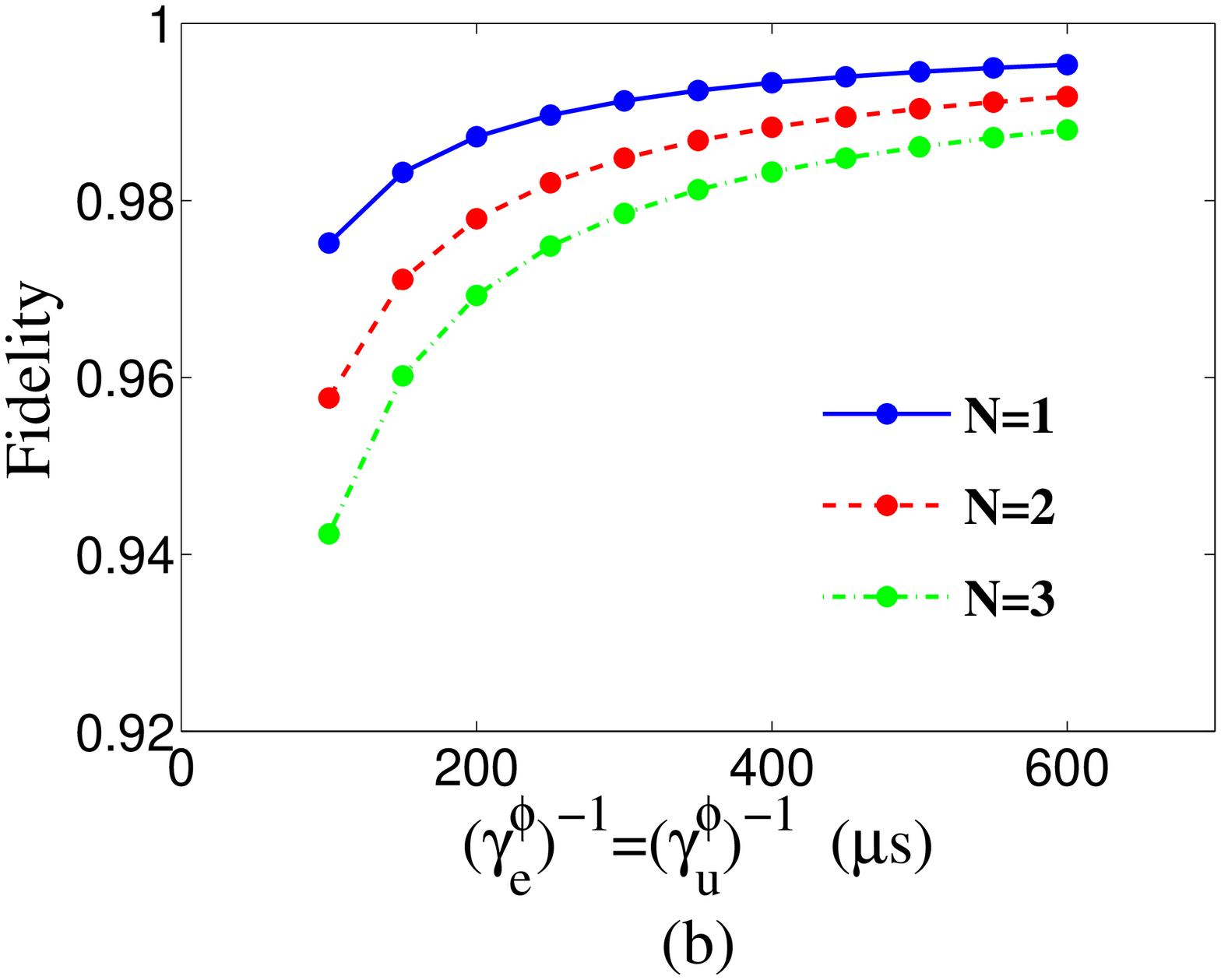}
\caption{(Color online) (a) The fidelities of the NOON states ($N=M=1$, $2$, and $3$) with three different couplings: $g_2^{ge}=g_2^{gu}=2\pi\times0.32$ MHz (blue square), $2\pi\times0.28$ MHz (red circle), and $2\pi\times0.35$ MHz (green triangle). (b) The fidelities of the NOON states vary with the dephasing rate of the NVE.}\label{fig2}
\end{figure}

To show the feasibility of the scheme, we numerically simulate \cite{NORI8,NORI9,Nathan} the fidelity of the NOON state through the Hamiltonian $H^r_j$ ($j=e, u, 1, 2$) by considering the relaxation rate and the dephasing rate of the NVE. The master equation governing the dynamics of the system is
\begin{eqnarray}        
\frac{d\rho}{dt} &=& -\frac{i}{\hbar}\left[H_j^r,\rho\right]+\kappa_1 D\left[b_1\right]\rho+\kappa_2 D\left[b_2\right]\rho +\gamma_{ge} D\left[\sigma_{ge}\right]\rho+\gamma_{gu} D\left[\sigma_{gu}\right]\rho \nonumber \\
&&+\gamma_e^\phi\left(\sigma_{ee}\rho\sigma_{ee}-\frac{1}{2}\sigma_{ee}\rho-\frac{1}{2}\rho\sigma_{ee}\right) +\gamma_u^\phi\left(\sigma_{uu}\rho\sigma_{uu}-\frac{1}{2}\sigma_{uu}\rho-\frac{1}{2}\rho\sigma_{uu}\right).
\end{eqnarray}
Here, $\kappa_{1(2)}$ is the decay rate of the AWR$_{1(2)}$. $\gamma_{ge} (\gamma_{gu})$ is the energy relaxation rate of the transition  $\left|g\right\rangle\leftrightarrow\left|e\right\rangle$ ($\left|g\right\rangle\leftrightarrow\left|u\right\rangle$) of the NVE, and $\gamma_e^\phi (\gamma_u^\phi)$ is the dephasing rate of the state $\left|e\right\rangle (\left|u\right\rangle)$. $\sigma_{ee}=\left|e\right\rangle\left\langle e\right|$ and $\sigma_{uu}=\left|u\right\rangle\left\langle u\right|$. $D\left[O\right]\rho=(2O\rho O^\dag-O^\dag O\rho-\rho O^\dag O)/2$ with $O=b_1$, $b_2$, $\sigma_{ge}$, and $\sigma_{gu}$.
The fidelity of the NOON state is defined as
\begin{equation}          
F=_f\!\!\left\langle\psi\right|\rho(t)\left|\psi\right\rangle_f.
\end{equation}
Here, $\rho(t)$ is the realistic density operator after the operations on the initial state $\left|\psi\right\rangle_I$ with the Hamiltonian $H^r_j$ and the decoherence of the system. $\left|\psi\right\rangle_f$ is the final state after the ideal operations on the initial state $\left|\psi\right\rangle_I$.

Here, the parameters of the system are taken as: $\omega_1=2\pi\times152.5$ MHz, $\omega_2=2\pi\times184.8$ MHz \cite{IEEE}, $\omega_e=2\pi\times2.7$ GHz, and $\omega_u=2\pi\times2.88$ GHz. The couplings are taken as $g_1^{ge}=g_1^{gu}=g_2^{ge}=g_2^{gu}=2\pi\times0.32$ MHz, which means the coupling strength between a single NV and the AWR is $g_s/2\pi\sim1$ kHz \cite{cs} when there are $10^5$ NV-centers in the ensemble.
$\kappa_1^{-1}=\kappa_2^{-1}\!=9.83\times10^2 \,s$ \cite{IEEE}, $\gamma_{ge}^{-1}=\gamma_{gu}^{-1}=6\,ms$, $\left(\gamma_e^\phi\right)^{-1}\!=\left(\gamma_u^\phi\right)^{-1}\!=600\,\mu s$ \cite{NVE}.
The remaining parameters in each step for generating the NOON states with $N=M=1$, $2$, and $3$ are shown in Table \ref{tab1}, Table \ref{tab2}, and Table \ref{tab3}, respectively. The fidelities of our NOON states plotted in Fig.\ref{fig2}(a) indicate the fidelities of the NOON states with $N=M=1$, $2$, and $3$ reach $F_1=99.54\%$, $F_2=99.18\%$, $F_3=98.80\%$, respectively.

To discuss the influences of the imperfect relationship among parameters, for simplicity, we consider two conditions: (1) coupling strengths between each transitions of the NVE and the two AWRs are not equal to each other; (2) the influences of different dephasing rates of the NVE. To consider the first condition, we take $g_1^{ge}=g_1^{gu}>g_2^{ge}=g_2^{gu}$ and $g_1^{ge}=g_1^{gu}<g_2^{ge}=g_2^{gu}$ by keeping $g_1^{ge}=g_1^{gu}=2\pi\times0.32$ MHz unchanged. When we take
$g_2^{ge}=g_2^{gu}=2\pi\times0.28$ MHz, the fidelities of the NOON states are $99.5\%$, $99.1\%$, and $98.77\%$ for $N=M=1$, $2$, and $3$, respectively, as shown in Fig.\ref{fig2}(a). When we take $g_2^{ge}=g_2^{gu}=2\pi\times0.35$ MHz, the fidelities of the NOON states for $N=M=1$, $2$, and $3$ are $99.5\%$, $99.2\%$, and $98.8\%$, respectively.
To show the influences of the dephasing rate of the NVE, we show that the fidelity of the state varies with $(\gamma_{e(u)}^{\phi})^{-1}$ as shown in Fig.\ref{fig2}(b) with $N=M=1$, $2$, and $3$.

\section{SUMMARY}

We propose a scheme to generate a NOON state using AWRs in a system consisting of two AWRs coupled to an NVE. With the resonant interactions between the AWR (microwave pulse) and the NVE, the numerical simulation shows that the fidelities of our NOON states can reach $99.54\%$ for $N=1$, $99.18\%$ for $N=2$, and $98.80\%$ for $N=3$ by considering decoherence of the system.

\section*{ACKNOWLEDGEMENTS}

M. Hua was supported by the National Natural Science
Foundation of China under Grants No. 11704281 and No. 11647042.

\end{document}